\documentclass{iau}
\usepackage{graphicx}
\usepackage{nicefrac}
\renewcommand{\vec}[1]{\mbox{\boldmath$#1$}}

\title[Experimental realization of dynamo action: present status and prospects] 
{Experimental realization of dynamo action: present status and prospects}

\author[A. Giesecke, F. Stefani, T. Gundrum, G. Gerbeth, C. Nore \& J. L{\'e}orat]   
{Andr{\'e} Giesecke$^1$,
Frank Stefani$^1$, Thomas Gundrum$^1$, Gunter Gerbeth$^1$, Caroline Nore$^2$ \and Jacques L{\'e}orat$^3$}

\affiliation{$^1$Helmholtz-Zentrum Dresden-Rossendorf\\ P.O.B. 510119,
D-01314, Dresden, Germany \\ email: {\tt a.giesecke@hzdr.de} \\[\affilskip]
$^2$Laboratoire d'Informatique pour la
      M\'ecanique et les Sciences de l'Ing\'enieur (LIMSI), CNRS, \\ BP 133,
      F-91403 Orsay cedex, France \\email: {\tt nore@limsi.fr}\\[\affilskip]
$^3$Observatoire de Paris-Meudon,\\ 
place Janssen, F-92195 Meudon, France \\email: {\tt jacques.leorat@obspm.fr}}

\pubyear{2012}
\volume{294}  
\pagerange{119--126}
\jname{Solar and Astrophysical Dynamos and Magnetic Activity}
\editors{A.G. Kosovichev, E.M. de Gouveia Dal Pino, \&  Y. Yan, eds.}
\begin{document}

\maketitle

\begin{abstract}

In the last decades, the experimental study of dynamo action
has made great progress. However, after the dynamo experiments in Karlsruhe and Riga, the
von-K{\'a}rm{\'a}n-Sodium (VKS) dynamo is only the third facility that
has been able to demonstrate fluid flow driven self-generation of magnetic
fields in a laboratory experiment. 
Further progress in the experimental examination of dynamo action is
expected from the planned precession driven dynamo experiment that
will be designed in the framework of the liquid sodium facility DRESDYN (DREsden
Sodium facility for DYNamo and thermohydraulic studies). 
 
In this paper, we briefly present numerical models of the VKS dynamo
that demonstrate the close relation between the axisymmetric field
observed in that experiment and the soft iron material used for the
flow driving impellers. 
We further show recent results of preparatory water experiments and
design studies related to the precession dynamo and delineate
the scientific prospects for the final set-up.

\end{abstract}

\firstsection 

%
%
\section{Introduction}

Liquid metal dynamo experiments present a complementary tool to gain
further insight into the working principles of astrophysical magnetic
field generation, for example by allowing a verification of scaling
laws obtained from numerical simulations or by providing measurements
in a level of detail that cannot be reached with astronomical
observations.  Looking at the typical laboratory scale
($\mathcal{L}\sim 1\mbox{ m}$), the fluid flow driven generation of
magnetic fields is a demanding task that requires typical flow
velocities of the order of $10 \mbox{ m/s}$ in order to cross the
dynamo threshold at all. So far, only three facilities have been able
to demonstrate fluid flow driven self-generation of magnetic fields
(\cite[Stefani \etal, 2008]{stefani_zamm08}).  The first confirmation
of fluid flow driven dynamo action under laboratory conditions
occurred nearly simultaneously at two distinct experiments conducted
in Riga (\cite[Gailitis \etal, 2000]{gailitis00}) and in Karlsruhe
(\cite[Stieglitz \& M{\"u}ller, 2001]{stieglitz00}).  The Karlsruhe
dynamo experiment essentially consisted of a cylindrical assembly of
helical guiding tubes thus roughly mimicking the (assumed) flow in the
Earth's liquid core.  The small scale helical structure of the flow
configuration was suitable for applying a two-scale separation and the
observed generation of a large scale magnetic field could be well
described using mean-field theory.

An explanation of induction action from a mean flow field was also
appropriate for the Riga dynamo (\cite[Gailitis \etal,
  2004]{gailitis04}).
The principle of this experiment is based on a
flow configuration proposed by \cite{ponomarenko}, who showed that an
infinite helical flow embedded in a stationary conductor can show
dynamo action at a rather low magnetic Reynolds number.  The
realization in the experiment is carried out in a tall cylinder with a
single propeller driving a flow along the axial direction.  In
contrast to the Karlsruhe dynamo which has been deconstructed a couple
of years ago, the Riga dynamo is still operating allowing further
investigations like, e.g., a transition to chaotic behavior that has
recently been found numerically by \cite[Stefani
  \etal~(2011)]{stefani11}.

\section{The von-K{\'a}rm{\'a}n-Sodium dynamo}

In the VKS dynamo experiment a flow of liquid sodium is driven by two
counter-rotating impellers that are located close to the end-caps of a
cylindrical vessel (see left panel in figure 1).  Dynamo action was
found at a surprising low magnetic Reynolds number of
${\rm{Rm}}\approx 32$. In dependence of the flow driving various
regimes could be explored which show different dynamical properties,
like bursts, oscillations or sudden field reversals (\cite[Monchaux
  \etal, 2007; Berhanu \etal, 2007]{monchaux07, berhanu07}).  A
striking property of the VKS dynamo is the exclusive occurrence of
(axisymmetric) dynamo action only in cases when the flow is driven by
impellers that are made of a soft iron alloy with a relative
permeability of $\mu_{\rm{r}}\approx 65$ (\cite[Monchaux \etal, 2009;
  Verhille \etal, 2010]{monchaux09, verhille10}).
\begin{figure}[h!]
\begin{center}
 \includegraphics[width=13cm]{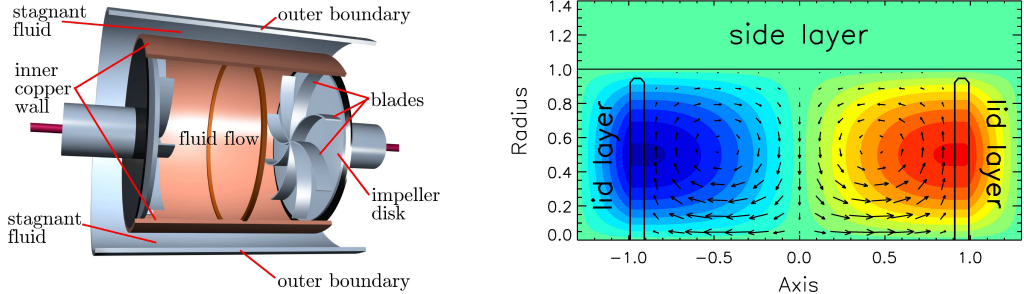} 
 \caption{Left: Sketch of the original setup of the VKS dynamo. Later,
   the inner copper walls have been removed so that the flow active
   regime was enlarged and the aspect ratio (height over diameter)
   decreased. Right: Mean velocity field
   applied in the numerical simulations. The color coded structure
   presents the azimuthal velocity and the arrows denote the
   meridional velocity contribution ($v_r, v_z$). The velocity field
   stems from analytical expressions specified by \cite{marie06}.   
   Two disk-like sub-domains with radius $R_{\rm{imp}}=0.95$ are located
   in the intervals $z\in [-1.0;-0.9]$ and $z\in [0.9;1.0]$ and
   represent soft iron disks of thickness $d=0.1$.}
   \label{fig1}
\end{center}
\end{figure}

\begin{figure}[h!]
  \begin{center}
    \includegraphics[width=10.5cm,draft=false]{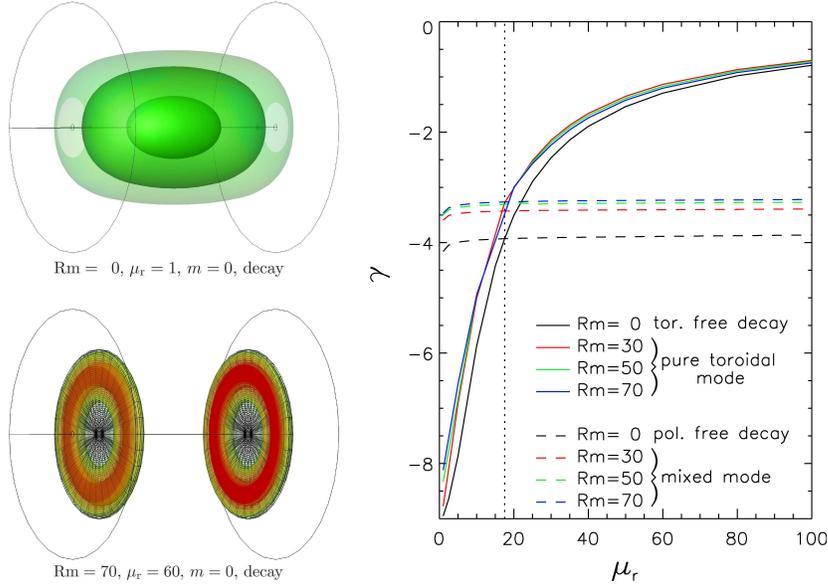}
    \caption{Field structure and growth-rates for the axisymmetric 
      eigenmodes. 
      The iso-surfaces on the left present the magnetic energy density of the $m=0$ mode for
      various sets of the control parameters ${\rm{Rm}}$ and $\mu_{\rm{r}}$. The
      plot on the right hand side shows the corresponding growth-rates of the
      $m=0$ eigenmode (solid curves: purely toroidal mode, dashed
      curve: mixed mode).}\label{fig2} 
  \end{center}
\end{figure}

In order to include the impact of a non-uniform distribution of
the relative permeability $\mu_{\rm{r}}$ the induction equation must
be written in the form  
\begin{equation}
\frac{\partial{\vec{B}}}{\partial t}=
\nabla\times\left(\vec{u}\times\vec{B}
+\frac{1}{\mu_{\rm{r}}\mu_0\sigma}\frac{\nabla\mu_{\rm{r}}}{\mu_{\rm{r}}}
\times\vec{B}-\frac{1}{\mu_{\rm{r}}\mu_0\sigma}\nabla\times\vec{B}\right)\label{eq1}
\end{equation}
where $\vec{u}$ is the prescribed (mean) flow, $\vec{B}$ the magnetic flux
density $\mu_0$ the vacuum permeability 
($\mu_0=4\pi\cdot 10^{-7}\mbox{ Vs/Am}$) and $\sigma$ the electrical conductivity.  
We have performed numerical simulations of (\ref{eq1}) using a
prescribed analytical flow field  which
resembles the average flow resulting from a von-K{\'a}rm{\'a}n-like
forcing (the so called MND-flow, see right panel of figure~\ref{fig1}
and \cite[Mari{\'e} \etal, 2006]{marie06}).  
The soft iron material of the impeller disks is modelled by
two axisymmetric permeability distributions. 
The results in terms of eigenmodes and related growth-rates reveal a
close connection between the exclusive occurrence of dynamo action in
the presence of soft-iron impellers and the observed axisymmetry of
the magnetic field (see figure~\ref{fig2} and 
\cite[Giesecke \etal, 2012]{giesecke12}).  

Here we consider only the behavior of the axisymmetric eigenmodes, for
which we find two distinct classes of eigenmodes.  A so called
{\it{mixed mode}}, consisting of a poloidal and a toroidal
contribution, is largely independent of the permeability. This mode
always decays on a rather fast timescale.  Furthermore, we observe a
purely toroidal mode that is considerably enhanced with increasing
$\mu_{\rm{r}}$. This mode essentially stems from the paramagnetic
pumping generated at the interface between fluid and soft-iron disk
(\cite[Giesecke \etal, 2012]{giesecke12}) and is largely independent
of the flow magnitude.  Increasing the disk permeability this mode is
shifted close to the dynamo threshold and becomes the leading
eigenmode of the system within the experimentally relevant regime
(i.e. for ${\rm{Rm}}\leq 50$ and $\mu_{\rm{r}}\approx 65$).  However,
regarding the actual axisymmetric setup the purely toroidal mode is
not able to become a growing eigenmode due to the restrictions
resulting from Cowling's theorem.  Hence, a satisfying explanation of
the observed axisymmetric dynamo mode requires mean field effects like
the $\alpha$-effect. Since the flow is considerably turbulent such
effects are undoubtedly operative, however, so far their properties
(e.g. spatial distribution or amplitude) are only speculative.  The
$\alpha$-effect is closely related to the kinetic helicity via the
well know relation
$\alpha\sim\nicefrac{\tau}{3}\left<\vec{u}\cdot(\nabla\times\vec{u})\right>$
(\cite[Krause \& R{\"a}dler, 1980]{krause80}). Very recently,
\cite[Ravelet \etal~(2012)]{ravelet12} presented a rough estimate for
the kinetic helicity
$h_{\rm{kin}}=\left<\vec{u}\cdot(\nabla\times\vec{u})\right>$ obtained
from numerical simulations which shows strong spatial concentrations
close to the impellers. The corresponding magnitude for $\alpha$ is
definitely within the regime that was required in \cite[Giesecke \etal
  (2010)]{giesecke10} to allow for growing axisymmetric dynamo modes
(using a uniform $\alpha$-distribution).

\section{A precession driven dynamo}

The DREsden Sodium facility for DYNamo and thermohydraulic studies
(DRESDYN) is a scheduled infrastructure project which will serve as a
platform for large-scale experiments related to geo-and astrophysics
as well as for thermohydraulic experiments on liquid metal
applications in energy related technologies (\cite[Stefani \etal,
  2012]{stefani12}).  The most elaborate facility in the framework of
DRESDYN will be a precession-driven dynamo experiment. Further planned
experiments are a large Taylor-Couette type experiment for the
combined investigation of the magneto-rotational instability and the
Tayler instability and various small scale experiments related to the
thermo-hydraulics of liquid sodium.  In the following, we will give a
brief description of the setup of the precession dynamo and a short
summary of preliminary results from a preparatory water experiment.

\subsection{Motivation and theoretical background}

Precession has often been regarded as an alternative or complementary
driving mechanism for the geodynamo in order to overcome
inconsistencies related to the age of the Earth's solid inner core
(\cite[Malkus, 1968]{malkus68}).  The precession of the Earth's axis
represents a significant variance of an orbital parameter which
undoubtedly influences the Earth's inner core flow and hence the
geodynamo.  This is supported by modulations of the inter-reversal
time distribution with a period $\tau\sim 100 {\mbox{ kyrs}}$ which is
rather close to the Milankovich cycle period that describes an
oscillation in the eccentricity of the Earth's orbit (\cite[Consolini
  \& De Michelis, 2003]{consolini03}).  A dynamo experiment with a
fluid flow driven by precession is further attractive because a
conducting fluid simultaneously rotating around two axis provides the
preconditions for ideal homogeneous dynamo action without any internal
guiding tubes (as in Karlsruhe) or propellers (as in Riga and the VKS
dynamo).

In a co-rotating system subject to precessional driving the flow is
determined by the Navier-Stokes equation including source terms for
the Coriolis- and the Poincar{\'e} forces:
\begin{equation}
\frac{\partial}{\partial
  t}\vec{u}+(\vec{u}\nabla)\vec{u}+2(\vec{\omega}+\vec{\Omega(t)})\times\vec{u}
=\nu\nabla^2\vec{u}-\nabla\varPhi
-(\vec{\Omega(t)}\times\vec{\omega})\times\vec{r}.\label{eq::navierstoke} 
\end{equation}
In eq.~(\ref{eq::navierstoke}) $\vec{u}$ is the velocity field,
$\vec{\omega}$ denotes the angular velocity of the fluid container,
$\vec{\Omega}$ is a time-dependent vector that describes the
precession of $\vec{\omega}$ and $\varPhi$ is the reduced pressure
(including the centrifugal term).  Numerical simulations in various
geometries have shown that a flow described by~(\ref{eq::navierstoke})
indeed is able to drive a dynamo (sphere, \cite[Tilgner,
  2005]{tilgner05}; cylinder, \cite[Nore \etal, 2011]{nore11}; cube,
\cite[Krauze, 2010]{krauze10}; spheroid, \cite[Wu \& Roberts,
  2009]{wu09}).  However, so far the experimental verification which
of course will be realized at parameters that will be quite different
from the numerical simulations is missing and reliable conclusions for
natural dynamos remain difficult.

\subsection{Experimental set-up}

\begin{figure}[h!]
\begin{center}
 \includegraphics[width=6cm,angle=-90]{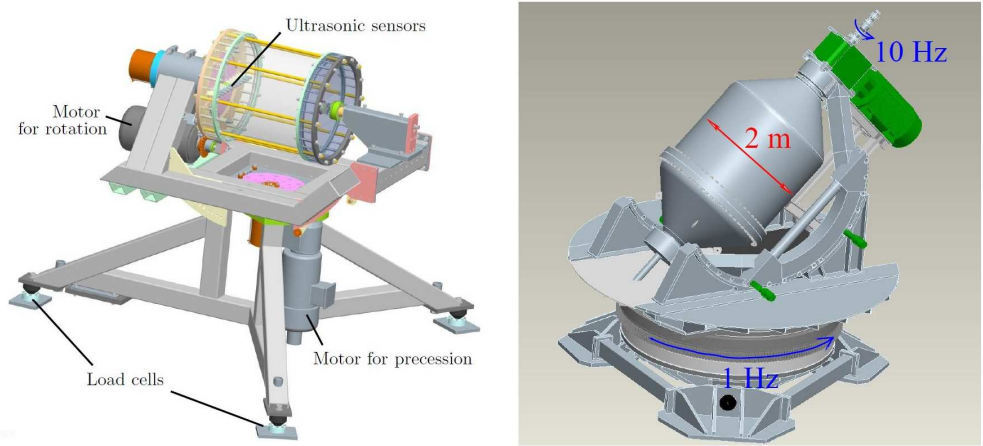} 
\caption{Left: Sketch of the small-scale water precession
  experiment (1:6 downscaled). The experiment provides the possibility to observe the
  velocity field (via Ultrasonic Doppler Velocimetry (UDV)). Right: Draft sketch for the
  large-scale precession dynamo experiment. The frequencies denote
  the maximum achievable frequencies for rotation and precession. The
  diameter of the inner cylinder will be approximately $2{\mbox m}$
  with an aspect ratio close to 1. \label{fig3}}
\end{center}
\end{figure}

The precession dynamo scheduled in the framework of DRESDYN will
consist of a cylindrical container with approximately $2\mbox{ m}$
diameter, rotating with up to $\omega_{\rm{cyl}}=10 \mbox{ Hz}$ around
its axis. The vessel will additionally rotate with up to
$\omega_{\rm{p}}=1 \mbox{ Hz}$ around a second axis, the precession
axis, whose angle with respect to the first axis can be varied between
$90^{\circ}$ and $45^{\circ}$ (see right panel in figure~\ref{fig3}
for a preliminary sketch).
In preparation for the liquid sodium experiment and in order to
determine the optimal geometric configuration as well as essential
process parameters a smaller water experiment has been developped (see
left panel in figure~\ref{fig3}).  This small scale experiment is
similar to the ATER experiment(\cite[L{\'e}orat, 2006; Mouhali,
  2010]{leorat06,mouhali2010}) but is equipped with additional sensors
that provide the determination of torques and motor powers needed to
drive the rotation of the cylinder and the turntable. Further problems
that are attacked in the water experiment are the estimation of the
gyroscopic torques acting on the basement and the estimation of the
average flow that can be applied for kinematic simulations.

\begin{figure}[h!]
\begin{center}
\includegraphics[width=8cm]{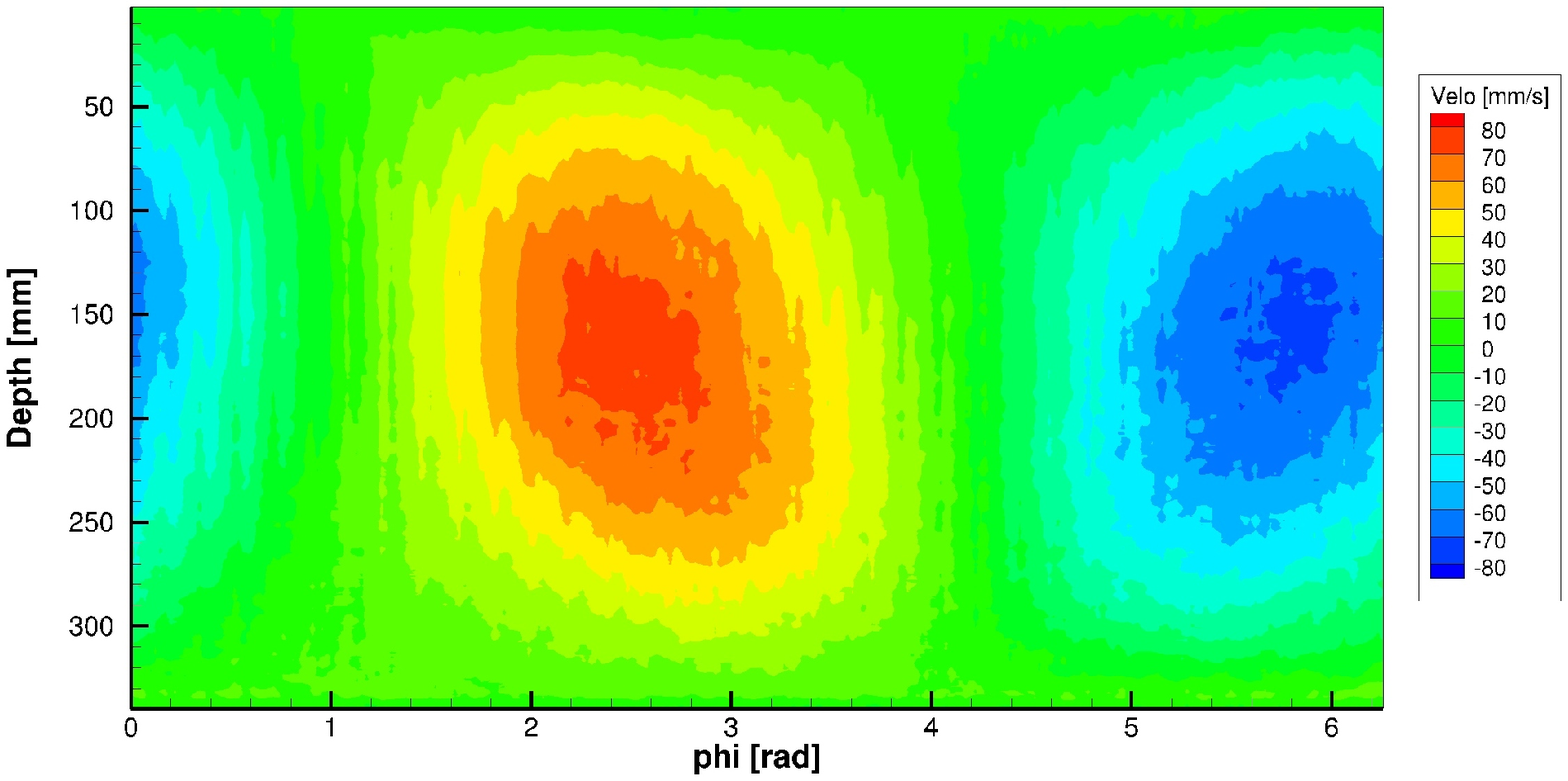} 
\nolinebreak[4!] 
\includegraphics[width=5.5cm]{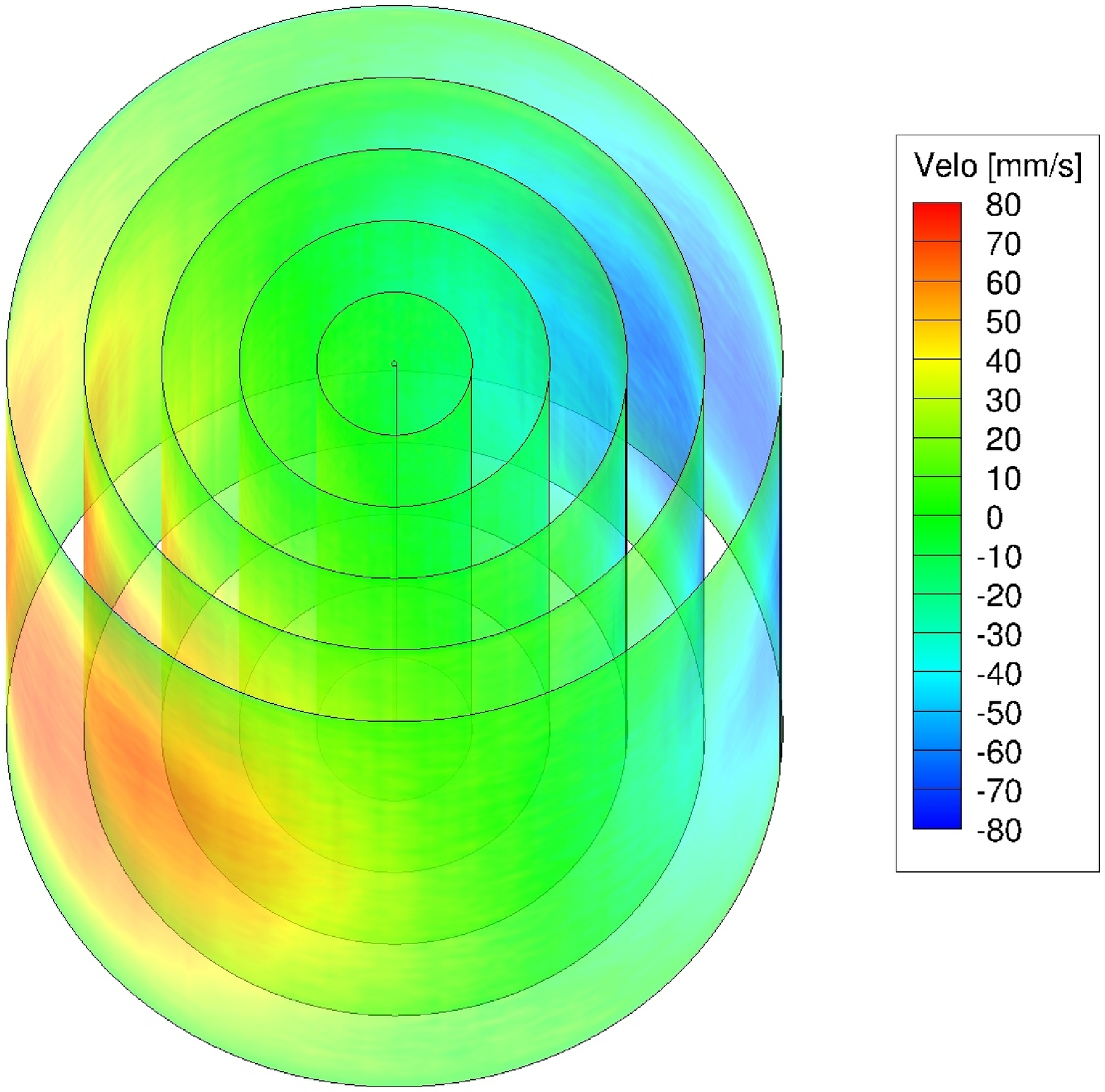} 
\caption{Pattern for a weakly precessing flow
  ($\omega_{\rm{cyl}}=0.2\mbox{ Hz}$ and $\omega_{\rm{p}}=0.01\mbox{
    Hz}$). The colors denote the axial velocity $u_z$ in the
  co-rotating frame. A clear non-axisymmetric mode with an azimuthal
  wavenumber $m=1$ has emerged. The maximum velocity magnitude is
  roughly one third of the rotation of the container.\label{fig4}}
\end{center}
\end{figure}

Up to present, various ultrasonic devices are installed on the end
caps of the cylinder allowing the determination of the axial velocity
component using {\it{Ultrasonic Doppler Velocimetry}} (UDV).  In the
slowly rotating regime and for low precession rates
$\Gamma=\omega_{\rm{p}}/\omega_{\rm{cyl}}$ we first observe a laminar
flow comprising only a few non-axisymmetric modes.  The large scale
flow component is determined by an azimuthal wavenumber $m=1$ and the
flow structure is typical for a {\it{Kelvin mode}} which is fixed in
the frame of the turntable (see figure~\ref{fig4}).
For $\omega_{\rm{cyl}}=0.2\mbox{ Hz}$ the typical velocity amplitudes
in the co-rotating system (i.e. the container-wall frame) are of the order of
$u_z\approx 40\mbox{ mm/s}$. 
Scaled to the 6 times larger liquid sodium facility and to a
rotation rate of $\omega_{\rm{cyl}}=10\mbox{ Hz}$ this would result in
a a value of $12\mbox{ m/s}$ which correponds to a magnetic Reynolds number of 
${\rm{Rm}}\sim 240$.

Increasing the precession rate above $\Gamma\approx 0.07$ the flow
abruptly switches into a fully turbulent state which is accompied by a
sharp increase of the required motor power.  So far, the fully
turbulent regime (as well as the transitional regime) cannot be
reached in numerical simulations. However, from the water experiment it is
already obvious that the flow properties in both regimes are different
with the simple $m=1$ mode being suppressed in the turbulent state
so that we also expect surprising effects for dynamo action, e.g. an increased
impact of the magnetic field on the fluid flow.

\section{Conclusions}

Fluid flow driven laboratory dynamo action is an established
phenomenon but still not easy to achieve. Further progress is expected
utilizing "natural" flows as energy source for a dynamo as it is
planned for the precession dynamo experiment in the framework of
DRESDYN.  For the slow rotation rate examined so far the flow
structure most probably is too simplistic to provide for dynamo
action.  Nevertheless, the preparatory water experiments show that
precessional flow driving is quite efficient and will allow to reach
magnetic Reynolds numbers that will be in the range of the critical
value required to achieve precessional driven dynamo action in a
sphere in the simulations of \cite{tilgner05}.  More complex flow
geometries are expected for higher precession rates and the next goal
for the flow measurements is an identification of helical eddies that
have been found by \cite{mouhali10} at the ATER experiment.


\begin{thebibliography}{}

\bibitem[Stefani \etal (2008)]{stefani_zamm08}
{{Stefani}, F., {Gailitis}, A. \& {Gerbeth}, G.} 2008,
\textit{Z. Angew. Math. Mech.}, 88 (12), 930--954

\bibitem[Berhanu \etal (2007)]{berhanu07}
{{Berhanu}, M. \etal} 2007,
\textit{Europhys. Lett.}, 77, 59001


\bibitem[Consolini \& De Michelis (2003)]{consolini03}
{{Consolini}, G. \& {De Michelis}, P.} 2003,
\textit{Phys. Rev. Lett.}, 90, 058501


\bibitem[Gailitis \etal (2000)]{gailitis00}
{Gailitis, A. \etal} 2000, 
\textit{Phys. Rev. Lett.}, 84 (19), 4365--4368 


\bibitem[Gailitis \etal (2004)]{gailitis04}
{{Gailitis}, A. \etal} 2004, 
\textit{Phys. Plasmas}, 11, 2838--2843 

\bibitem[Giesecke \etal (2010)]{giesecke10}
{{Giesecke} A., {Stefani}, F. \& {Gerbeth}, G.} 2010, 
\textit{Phys. Rev. Lett.}, 104, 044503


\bibitem[Giesecke \etal (2012)]{giesecke12}
{{Giesecke}, A. \etal} 2012,  
\textit{New J. Phys.}, 14 (5), 053005

\bibitem[Krauze (2010)]{krauze10}
{{Krauze}, A.} 2010, 
\textit{Magnetohydrodynamics}, 46 (3), 271--280 

\bibitem[Krause \& R{\"a}dler (1980)]{krause80}
{{Krause}, F. \& {R{\"a}dler}, K.-H.}, 
\textit{Mean-field magnetohydrodynamics and dynamo theory}, Oxford,
Pergamon Press, 1980 

\bibitem[L{\'e}orat (2006)]{leorat06}
{{L{\'e}orat}, J.} 2006, 
\textit{Magnetohydrodynamics}, 42 (2--3), 143--151 

\bibitem[Malkus (1968)]{malkus68}
{{Malkus}, W.V.R.} 1968, 
\textit{Science}, 160, 259--264

\bibitem[Mari{\'e} \etal (2006)]{marie06}
{{Mari{\'e}} L., {Normand} C. \& {Daviaud} F.} 2006,
\textit{Phys. Fluids}, 18, 017102


\bibitem[Monchaux \etal (2007)]{monchaux07}
{{Monchaux}, R. \etal} 2007, 
\textit{Phys. Rev. Lett.}, 98, 044502


\bibitem[Monchaux \etal (2009)]{monchaux09}
{{Monchaux}, R. \etal} 2009
\textit{Phys. Fluids}, 21 (3), 035108

\bibitem[Mouhali (2010)]{mouhali10}
{{Mouhali}, W.} 2010, 
\textit{PhD Thesis}, Universit{\'e} Paris-Diderot -- Paris VII  

\bibitem[Nore \etal (2011)]{nore11}
{{Nore}, C., {L{\'e}orat}, J., {Guermond}, J.-L. \& {Luddens}, F.} 2011, 
\textit{Phys. Rev. E}, 84 (1), 016317

\bibitem[Stefani \etal (2008)]{stefani08}
{{Stefani}, F., {Gailitis}, A. \& {Gerbeth}, G.} 2008, 
\textit{Z. Angew. Math. Mech.}, 88, 930--954 

\bibitem[Stefani \etal (2011)]{stefani11}
{{Stefani}, F., {Gailitis}, A. \& {Gerbeth}, G.} 2011, 
\textit{Astron. Nachr.}, 332, 4


\bibitem[Stefani \etal (2012)]{stefani12}
{{Stefani}, F. \etal} 2012,
\textit{Magnetohydrodynamics}, 48 (1), 103--113 

\bibitem[Stieglitz \& M{\"u}ller (2001)]{stieglitz00}
{{Stieglitz}, R. \& {M{\"u}ller}, U.} 2001, 
\textit{Phys. Fluids}, 13, 561--564

\bibitem[Tilgner (2005)]{tilgner05}
{Tilgner, A.} 2005,
\textit{Phys. Fluids}, 17 (3), 034104 

\bibitem[Ponomarenko (1973)]{ponomarenko}
{{Ponomarenko}, Y.-B.} 1973, 
\textit{J. App. Mech. Tech. Phys.}, 14, 775--778

\bibitem[Ravelet \etal (2012)]{ravelet12}
{{Ravelet}, F., {Dubrulle}, B., {Daviaud}, F. \& {Rati{\'e}}, P.-A.} 2012
\textit{Phys. Rev. Lett.}, 109 (2), 024503


\bibitem[Verhille \etal (2010)]{verhille10}
{{Verhille}, G. \etal} 2010 
\textit{New J. Phys.}, 12 (3), 033006

\bibitem[Wu \& Roberts (2009)]{wu09}
{{Wu}, C.-C \& {Roberts}, P.} 2009,
\textit{Geophys. Astrophys. Fluid Dyn.}, 103 (6), 467--501




\end{thebibliography}
\end{document}